\documentclass[a4paper,11pt]{article}
\pdfoutput=1 

\usepackage{jcappub} 
\usepackage[numbers]{natbib}
\usepackage[T1]{fontenc} 
\usepackage[table]{xcolor}
\usepackage[utf8]{inputenc}
\usepackage[english]{babel}

\title{\boldmath Effect of the brightest gamma-ray burst (GRB 221009A) on low energy gamma-ray counts at sea level}

\author[a]{Pranali Thakur,}
\author[a]{Gauri Datar,}
\author[a]{Geeta Vichare,}
\author[b]{Selvaraj Chelliah}

\affiliation[a]{Indian Institute of Geomagnetism (IIG), Sector 18, Navi Mumbai, India}
\affiliation[b]{Equatorial Geophysical Research Laboratory, IIG, Tirunelveli, India}

\emailAdd{pranalispace@gmail.com}
\emailAdd{datar.gouri@gmail.com}
\emailAdd{vicharegeeta@gmail.com}
\emailAdd{selvaegrl@gmail.com}

\abstract{A gamma-ray burst, named GRB 221009A, occurred on 9 October 2022 and is the brightest ever observed GRB, whose frequency is now estimated as once in 10,000 years. This GRB was reported to be observed from many space missions, VLF receivers, and ground observations in optical and radio data. Additionally, a strikingly large number of very high energy (VHE) photons associated with this GRB were observed by the gamma-ray and cosmic ray observatory LHAASO. Though gamma-rays of cosmic origin usually tend to be absorbed by the atmosphere, the high fluence of this GRB, along with the observation of more than 5000 VHE photons (0.5 to 18 TeV) by LHAASO from the ground, emphasises the need to explore other possible ground observations of this GRB.

With RA = 288.3$^{o}$ and Dec = 19.8$^{o}$, the exceptionally bright fluence of this GRB was geographically centred on India. The present paper examines the effect of this GRB using gamma-ray data in a low energy range (0.2 -- 6) MeV obtained using NaI (Tl) detectors located at Tirunelveli (Geographic coordinates: 8.71$^{o}$N, 77.76$^{o}$E), India. We report no significant change in the observations associated with GRB 221009A. We discuss the extent of attenuation of gamma-rays in the atmosphere that could explain the reported observations. Further, we investigate the likelihood of ground observation of gamma-rays ($<$10 MeV) for a much more intense hypothetical GRB and estimate the parameters, such as distance, fluence, and isotropic energy of such a GRB.}


\begin{document}
\maketitle
\flushbottom
\section{Introduction}
\label{sec:intro}

Gamma-ray bursts (GRBs) are bright flashes of radiation with spectral energy distributions peaking in the gamma-ray band \citep{gehrels2009gamma}. Since their discovery, GRBs have been studied extensively from various aspects as they provide a natural laboratory of extreme physics. Apart from astrophysical studies in the standard domain, there are studies reporting on possible terrestrial impacts of GRBs \citep{melott2004did,melott2005climatic,thomas2005terrestrial,thomas2005gamma}. They have speculated that nearby and very intense GRBs can be the cause of extinction-like events in Earth's history, such as the Ordovician-Silurian mass extinction (445-415 million years ago) is hypothesized to have occurred due to a GRB at distance 2 $~kpc$ with fluence 100 $kJ/m^2$ \citep{thomas2005terrestrial}. Because of the possibility of such extreme effects, a recent GRB 221009A called the brightest of all time, or BOAT \citep{burns2023grb}, needs an investigation of its effects on the ground. GRB 221009A received a lot of attention in the scientific community. This specific event was observed from beyond 10 TeV by Large High Altitude Air Shower Observatory (LHAASO) \citep{huang2023invisible,lhaaso2023tera} down to radio frequencies \citep{fulton2023optical}. The optical data provided evidence of this GRB being associated with a supernova \citep{fulton2023optical}. 

 Many space-based missions detected the extremely intense GRB 221009A: Fermi-GBM (\citep{lesage2023fermi,veres2022grb}, Fermi-LAT \citep{bissaldi2022grb}), Konus-Wind (KW) \citep{frederiks2023properties}, AGILE \citep{piano2022grb,ursi2022grb}, Integral \citep{rodi2023soft}, Insight-HXMT and GECAM-C \citep{an2023insight}, Solar Orbiter \citep{xiao2022grb221009a}, SRG/ART-XC \citep{frederiks2023properties}, BepiColombo \citep{kozyrev2022improved}, CSES-01 \citep{battiston2023observation}, etc. The prompt emission started at 13:16:59 UT and lasted for $\sim$ 10 s initial pulse (IP), and the second pulse (P1) was observed after $\sim$ 180 s since trigger time, which lasted for 100 s \citep{veres2022grb}. The burst prompt emission had a complex time profile consisting of two distinct emission episodes. From a detailed analysis of KW G2 (80 -- 320) keV and ART-XC (in orbit at L1 and L2, respectively), Frederiks et al. (2023) \citep{frederiks2023properties} reconstructed this complex profile. According to their analysis, it starts with a single IP, which is followed, after a period of quiescence, by an extremely bright emission complex that lasts for $\sim$ 450 s and shows four prominent peaks: P1, at the onset P1 (at 13:19:55 UT), two huge pulses P2 and P3, a much longer but less intense P4. After $\sim$ 600 s, the prompt, pulsed phase of the burst evolves to a steadily decaying, extended emission tail, which is visible in the KW data for more than 25 ks \citep{frederiks2023properties}.  

The redshift of this GRB is estimated to be z $=$ 0.151 based on optical observations by \citep{de2022grb,malesani2023brightest}. A significant isotropic equivalent energy $E_{iso}$ in the 1-10000 keV range was estimated as (1.01 $\pm$ 0.007) $\times$ 10$^{55}$ erg from the redshift, and the fluence was measured as $0.21~erg ~cm^{-2}$ by KW \citep{frederiks2023properties} and $\sim$ $0.19~erg~cm^{-2}$ by Fermi-GBM \citep{lesage2023fermi}. The afterglow of GRB 221009A is also observed by the James Webb Space Telescope (JWST) and Hubble Space Telescope (HST) \citep{levan2023first}.
The optical observation of late-time flattening in the afterglow decay of this GRB is reported by \citep{srinivasaragavan2023sensitive} using GROWTH-India Telescope (GIT) \citep{kumar2022grb}, Lowell Discovery Telescope (LDT) \citep{o2022grb220611a,o2023structured}, and the Gemini-South Telescope (GST) \citep{rastinejad2022grb221009a}.

Another important observation was from LHAASO, a 1 km$^2$ air shower array at an altitude of 4,410 m a.s.l. in China. Their Water Cherenkov Detector Array (WCDA) detected more than 64000 photons associated with GRB 221009A, in 3000 s from the trigger. These photons were of energies between $\sim$ 200 GeV and $\sim$ 7 TeV. LHAASO observed these photons for about 6000 seconds until they moved out of the field of view. Also, Carpet-2 at Baksan Neutrino Observatory (1,700 m a.s.l.) observed an air shower caused by a photon of energy 251 TeV \citep{dzhappuev2022swift}. LHAASO Collaboration \citep{lhaaso2023tera} suggests that the TeV emission has a different origin than the prompt MeV emission.

This intense GRB ionised the upper parts of Earth's atmosphere. A sudden disturbance in the propagation of radio or very low frequency (VLF) band between 18 kHz and 23 kHz, i.e., sudden ionospheric disturbance (SID), was observed by various Indian lightning detectors coincident with the GRB 221009A \citep{guha2022grb221009a}. Similar observations were also reported by VLF-Monitor at Todmorden (UK) and Kiel Longwave Monitor (Germany) \citep{pal2023first}. 

The on-ground calculated location, using the GBM trigger data, was RA $=$ 288.3$^o$, DEC $=$ 19.8$^o$ (J2000 degree, equivalent to 19h22m, 22d5m) \citep{lesage2023fermi}. This location was at the zenith as seen from India at the time of the occurrence of GRB \citep{battiston2023observation}. Hence, the observations from the Indian subcontinent are especially interesting. In this paper, we investigate if gamma-rays from the GRB 221009A reach the ground by examining the gamma-ray variation during the GRB. Data on the lower energy component of gamma-rays are from a gamma-ray detector set-up consisting of two NaI (Tl) scintillation detectors at the Equatorial Secondary Cosmic Ray Observatory (ESCRO), Tirunelveli. It is well known that primary photons create a number of secondary electrons, positrons, and photons (electromagnetic shower) while interacting with the Earth’s atmosphere, and well established by theoretical calculation \citep{carlson1937multiplicative,bhabha1937passage}, observational \citep{takada2011observation,errando2023detect}, and simulation \citep{hillas1996differences,sinnis2009air,volk2012tev} studies. Although generally gamma-rays of lower energies (0.1 -- 10 MeV) are absorbed in the atmosphere, it will be interesting to see if such intense emissions of gamma-rays differ from the usual scenario and how deep they can penetrate inside the atmosphere. Further, it will be relevant to compare this GRB with a hypothetical GRB of very large fluence so as to shed some light on the implications of such GRBs hypothesised in \cite{melott2004did,melott2005climatic,thomas2005terrestrial,thomas2005gamma}.


\section{Data and Methods}

The experimental set-up consists of two NaI (Tl) detectors that continuously record data with a time resolution of one minute. The experimental set-up is located at the Equatorial Geophysical Research Laboratory (EGRL), Tirunelveli (Geographic coordinates: 8.71$^{o}$N, 77.76$^{o}$E). The detectors record gamma-ray flux in the 100 keV -- 6 MeV energy range. Both the NaI (Tl) detectors have the same size crystals ($10.16~cm \times 10.16~cm \times 40.64~cm$), and their shape is rectangular cuboid. To make a distinction, hereafter, these detectors are denoted as detector 0 and detector 1. 

Gamma-rays measured on the ground originate from various sources, such as atmospheric secondary cosmic rays (SCR), terrestrial radioactivity, man-made concrete buildings, nuclear plants, etc. Of these, the SCR contributes considerably to the background gamma-ray population. When the primary cosmic rays (CR), mostly consisting of protons, arrive at Earth, below $\sim$ 35 km, the air becomes dense, and they collide with the atmospheric nuclei as they travel down. In each collision, a primary CR produces a large number of charged and neutral pions, kaons, protons, neutrons, etc., which all interact or decay further down and give rise to a nuclear cascade. Neutral pions produced in various interactions almost instantaneously decay into two gamma-rays, which may undergo pair production, producing an electron-positron pair, which can generate two gamma-rays through annihilation, while electrons and positrons can individually produce more gamma-rays via bremsstrahlung. Thus, by the time the shower reaches the ground, the single primary would have multiplied into thousands to billions of particles, depending on its energy. Apart from such gamma-rays of SCR origin, terrestrial radioactivity is another constant source of radiation. The radiation from radionuclides in the Earth's crust and their decay products contribute significantly to the gamma-ray flux measured on the ground, particularly below 3 MeV. In the energy spectrum of NaI (Tl) detector data, the following background radioactivity peaks are clearly identifiable: $^{214}$Bi (0.609 MeV), $^{208}$Tl (0.908 MeV), $^{214}$Bi (0.112 MeV), $^{40}$K (1.46 MeV), $^{214}$Bi (1.76 MeV), $^{206}$Tl (2.615 MeV), etc. Thus, the terrestrial radioactivity and SCR always contribute to the gamma-ray flux measured at the ground. More comprehensive details of the detector set-up and gamma-ray measurement near sea level can be found in \citep{vichare2018equatorial, datar2020causes, datar2020response, datar2021barometric}.

In the daily temporal profile of total gamma-ray counts of all energies detectable by the NaI (Tl) detector, a distinct diurnal pattern is present \citep{datar2020causes}. The pattern arises due to gamma-rays of terrestrial radioactivity origin; hence, it is observed only below 3 MeV. By comparing the GRB event day data with quiet days, we can eliminate this effect and consider the whole energy range available for observations on GRB day. The quiet days were selected from previous and following adjacent weeks by comparing gamma-ray data with the atmospheric parameters (temperature, pressure, humidity, rainfall, wind, etc.) obtained from an in-house Automatic Weather Station (AWS). These days were cross-checked with the interplanetary data parameters, Sym-H index from the OMNI database and  WDC Kyoto, to eliminate any effect of space weather phenomena.     

To determine the significance of NaI (Tl) data, we have calculated measurement error. For a single measurement of gamma-ray counts, $n$, the standard deviation in $n$ is calculated as $\sigma$ = $\sqrt{n}$. Assuming Gaussian distribution for large $n$, there is a 90$\%$ probability that $n$ $\pm$ 1.64$\sigma$ will contain the true value of $n$ \citep{knoll2010radiation, datar2020causes}.


\begin{figure*}[b!]
	\includegraphics[width=0.9 \textwidth]{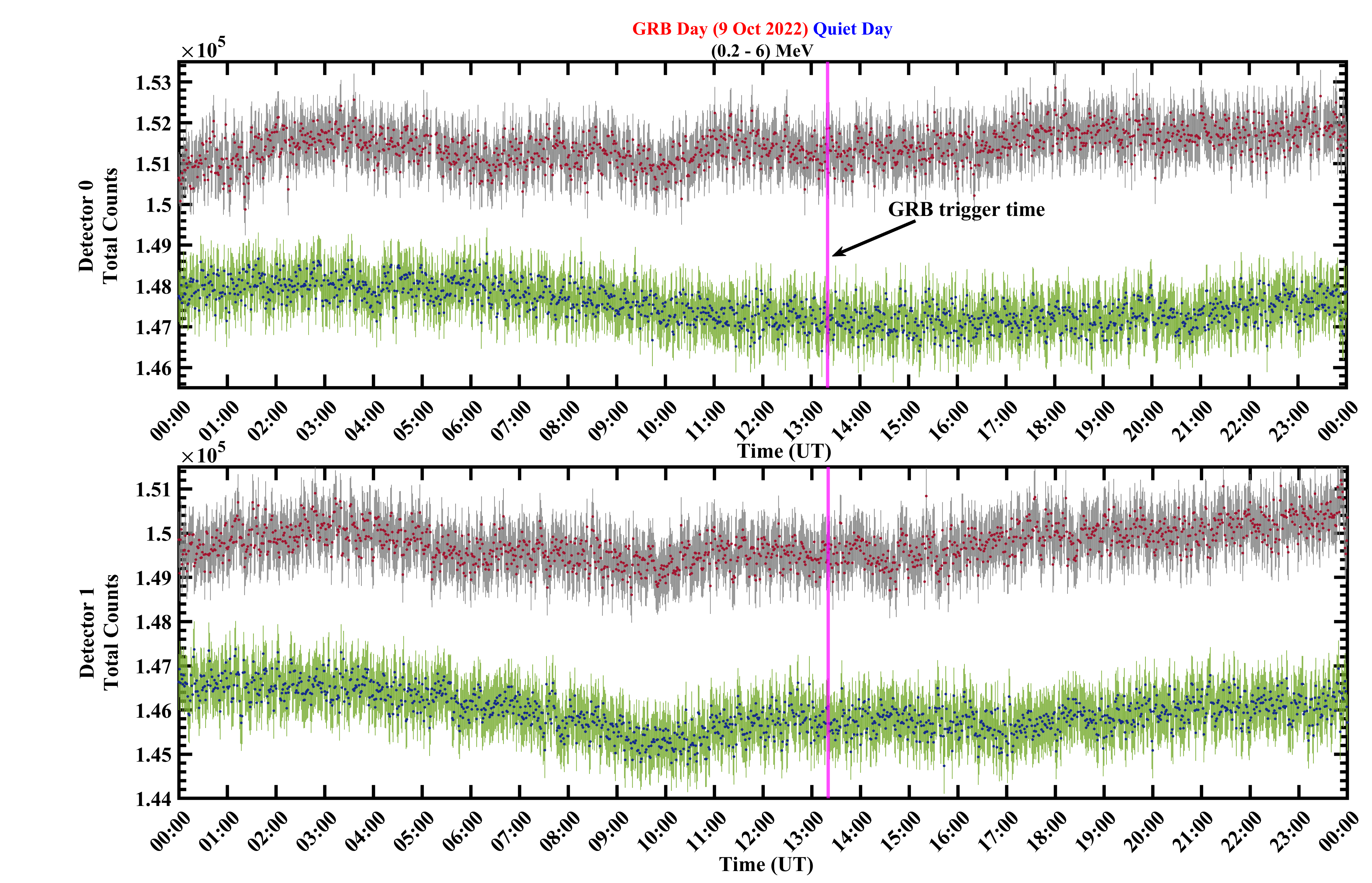}

	\caption{Total gamma-ray counts for (a) Detector 0 and (b) Detector 1, with the red-coloured dots showing the event day, whereas the blue-coloured dots are of the quiet day; the error bars of $\pm$ 1.64$\sigma$ are shown by grey and green for the event and quiet day, respectively. The vertical solid magenta line indicates the trigger time of the GRB event (13:20 UT).}
	\label{Figure1}
\end{figure*} 

\section{Observations}

\begin{figure*} [h]
\includegraphics[width=0.9 \textwidth]{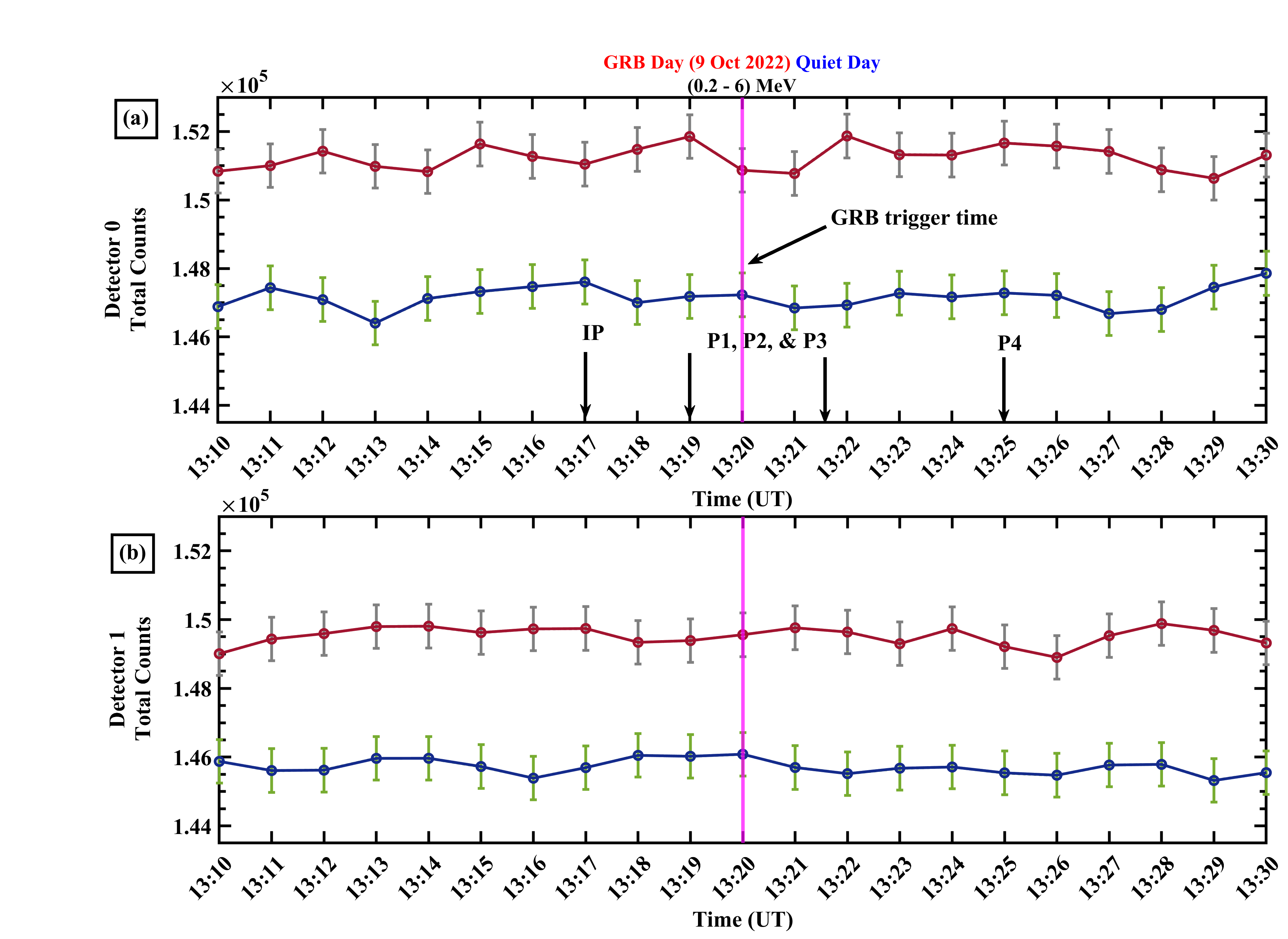}
\caption{Same as Figure \ref{Figure1}, but during a short observation window from 13:10 UT to 13:30 UT. Labels indicate the five peaks in GRB flux as discussed in Section 1 of this paper and Section 3 of \citep{frederiks2023properties}.}
\label{Figure2}
\end{figure*}

\begin{figure*} [h]
\includegraphics[width=1 \textwidth]{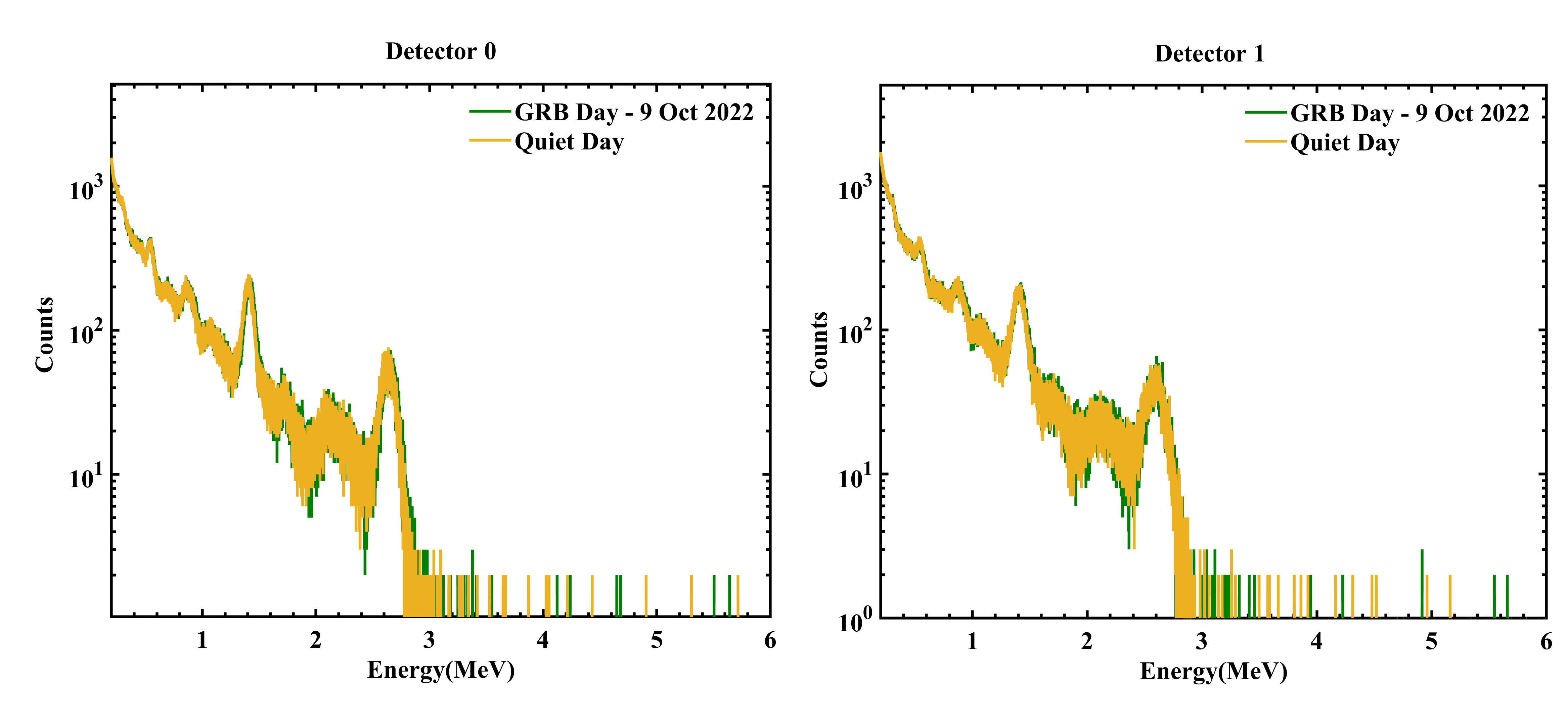}
\caption{Background gamma-ray spectra for Detector 0 (left) and for Detector 1 (right) were obtained by accumulating 11 minutes of data (13:15 to 13:26 UT) of GRB day and quiet day. Peaks are associated with gamma-ray emission lines associated with terrestrial radioactivity.}
\label{fig:FIG:spectra}
\end{figure*}

Figure \ref{Figure1}  presents the gamma-ray data in terms of total counts for (a)  detector 0 (top) and (b) detector 1 (bottom). In both the panels, the ${x}$-axis denotes time in UT, and the vertical magenta line indicates the trigger time of the GRB event (13:20 UT); the red-coloured dots show the event day, 9 October 2022, whereas the blue-coloured dots are of the quiet day (16 September 2022 and 14 October 2022 for detector 0 and detector 1, respectively), presented with vertical offset for clear distinction. The error bars are grey and green for the event and quiet day. The error is 1.64$\sigma$, as discussed in the previous section. For example, for a single measurement of counts of 1.51$\times$$10^5$, $\sigma$ = $\sqrt{n}$ = 388, and there is 90$\%$ probability that the true value is within n $\pm$ 1.64$\sigma$, i.e., 1.51$\times$ $10^5$ $\pm$ 636. Similarly, measurement error is calculated for all other observations, ensuring that all measurements are statistically 90$\%$ significant.

Figure \ref{Figure2} is a zoomed-in version of Figure \ref{Figure1}, centred on GRB occurrence time, and presents a short observation window from 13:10 UT to 13:30 UT, with the vertical magenta line indicating the trigger time of the GRB. The IP, P1, P2, P3, and P4 labels indicate the initial pulse and prominent peaks, respectively, as mentioned in \citep{frederiks2023properties}. 

Figures \ref{Figure1} and \ref{Figure2} show no significant difference in gamma-ray counts after the GRB trigger time. The slight variations are well within the range of statistical fluctuations. Thus, no significant change in counts over the quiet days is observed in the event day data for both detectors. 

To investigate further, Figure \ref{fig:FIG:spectra} compares the energy spectra of event days with quiet days. These spectra are each accumulated for 11 minutes on the event as well as quiet days, starting from 13:15 UT to 13:26 UT so as to encompass all the pulses of the prompt phase. The left panel in Figure \ref{fig:FIG:spectra} represents spectra for detector 0 and the right side for detector 1. The green-coloured data in both panels are of the event day, 9 October 2022, whereas the mustard-coloured data is of the quiet day. The spectra of both event and quiet days overlap in an exact manner, indicating no difference during the event time. 

\section{Discussion}

The gamma-ray counts in low energy (0.2 -- 6) MeV from NaI (Tl) detectors on the ground were studied during the GRB 221009A. From Figures \ref{Figure1}, \ref{Figure2}, and \ref{fig:FIG:spectra}, it is evident that no significant change was observed in gamma-ray counts of (0.2 -- 6) MeV energy range at our location. However, it has been reported previously that the cosmic ray and air shower observatory LHAASO has detected VHE photons associated with this GRB \cite{lhaaso2023tera}. Although the observation location in the present paper is in the geographical vicinity of LHAASO, there is a considerable difference in their altitudes. LHAASO is situated at 4,410 m a.s.l., whereas our location is at $\sim$ 30 m a.s.l. Another vital difference is the type of detectors and the energies they are capable of detecting. LHAASO's water Cherenkov detector array (WCDA) detected more than 64000 photons (200 GeV -- 7 TeV) within the first 3000 s of the GRB trigger \citep{lhaaso2023tera}. 

A large geographical region of this exceptionally large fluence GRB was centred on India, with a central impact point at 19.8$^o$ and 71$^o$ in latitude and longitude, respectively, mentioned in Figure 3 of \citep{battiston2023observation}. Therefore, it is interesting to examine the effect of this particular GRB using our gamma-ray detector data. Further, the GRB 221009A is known to have the highest recorded fluence in the known history of GRB observations. The fluence of this GRB is calculated as $0.21~erg~cm^{-2}$, whereas the second brightest GRB on record has a fluence of 4.56 $\times$ 10$^{-3}~erg~cm^{-2}$ (GRB 230307A) \cite{burns2023grb}. Thus, it is inquisitive to understand why gamma-rays ($<$10 MeV) are not observed on the ground during a GRB of the highest recorded fluence.

When a GRB occurs, photons of various energies are found in various observational domains: low energy ($<$10 MeV) photons are found in abundance at low-earth orbit (LEO, $\sim$ 500 km) or any other orbit outside the atmosphere, and a significant portion of very high energy (of GeV-TeV order) photons can manage to penetrate the atmospheric column and initiate Cherenkov air showers that Cherenkov detectors can detect. In the case of energies $<$500 GeV, the photon signal attenuates before it reaches the ground because of scattering from aerosols or high-altitude clouds. Such atmospheric scattering only weakly influences the propagation of very high energy ($>$1 TeV) gamma-rays \citep{snabre1998atmospheric}.

Once the GRB photons enter the atmosphere, they interact via various processes. Although Compton scattering dominates energy deposition in the (0.1 -- 10) MeV range, pair production dominates as the energy loss mechanism above 10 MeV \citep{das2003introduction}. When a high-energy photon ($>$100 MeV) undergoes pair production, an electron-positron pair is generated. Electrons and/or positrons can, in turn, again give rise to photons via annihilation or bremsstrahlung, and these photons will have a reduced energy than the original photon. Such secondary photons as a proxy of original photons may still be found on the ground, having penetrated the entire depth of the atmosphere. However, from the observations presented in the paper, it is likely that such secondary photons of energies below 10 MeV originating from higher energy photons could not reach the ground. 

As described in `Data and Methods', the NaI (Tl) detects photons of energies $<$10 MeV, with a significant contribution from terrestrial radioactivity and SCR, and in theory, is capable of detecting photons of GRB origin. However, the attenuation process affects photons and prevents the lower energy photons from reaching the ground, which is probably the case of photons of GRB origin. All photons, irrespective of their origin, undergo attenuation in the atmosphere. Nevertheless, photons produced as SCR via air shower, are generated recurringly at different levels/altitudes and by different particles in the numerous reactions of the cascade; thus, despite the attenuation, their penetration depths are different, making their observation on the ground feasible. GRB photons propagate through the atmosphere and are probably attenuated to such an extent as not to make their observation from the ground possible.

In this section, the altitude up to which the photons penetrated the atmosphere during GRB 221009A is estimated heuristically. Photons interact with matter through the photoelectric effect, Compton scattering, and pair production, and as a result, photon flux is attenuated exponentially after interaction with matter \citep{knoll2010radiation}. The total attenuation coefficient $\mu$ can be expressed as the sum of the individual coefficients due to each process \citep{das2003introduction}.

\begin{equation} \tag{1}\label{eq1}
	\mu = {{\mu}_{pe}} + {{\mu}_{c}} + {{\mu}_{pair}},
\end{equation}
where, $\mu_{pe}$, $\mu_{c}$, $\mu_{pair}$, are attenuation coefficients due to the photoelectric effect, Compton scattering, and pair production, respectively.
The flux intensity after travelling distance $x$ is measured by the Beer-Lambert attenuation equation \citep{lal1991cosmic,64354},
\begin{equation} \tag{2}\label{eq2}
	I = I{_{o}} . e^{\displaystyle{-\mu}{x}},
\end{equation}
where, $I{_{o}}$ is the intensity or photon counts at the beginning, $\mu$ is the linear attenuation coefficient $(m^{-1})$, and $x$ is the distance in metre $(m)$. If equation \ref{eq2} is modified as
\begin{equation} \tag{3}\label{eq3}
    	I = I{_{o}}.e^{\displaystyle-(\frac{\mu}{\rho}).\rho . x},
\end{equation}
where, $\rho~(kg/m^{3})$ is air mass density, then $\frac{\mu}{\rho}$ $(m^{2}/kg)$ is the mass attenuation coefficient, the values of which are available at \url{https://rais.ornl.gov/tools/Lin_Coefficients.pdf}. We have taken the height profile of the total air mass density from the \citep{chandrasekar2022basics} (US standard atmosphere).

Various LEO satellites like Fermi-(GBM \cite{veres2022grb} $\&$ LAT \citep{bissaldi2022grb}), AGILE \citep{ursi2022grb}, CSES-01 \citep{battiston2023observation}, SATech 01 \citep{liu2022grb}, etc., record photon counts in different energy ranges. From equation \ref{eq3}, and using the satellite measurements of counts during GRB 221009A as $I{_{o}}$, we estimate the propagation of gamma photons in the atmospheric column from 500 km to the ground. We calculated how photon counts of GRB 221009A attenuate with the depth travelled in the atmosphere, from LEO to Earth's surface (Table \ref{tab:Table1}). Here, photon counts from Fermi-GBM (\url{https://heasarc.gsfc.nasa.gov/FTP/fermi/data/gbm/triggers/2022/bn221009553/quicklook/}), SATech 01-HEBS (\url{http://twiki.ihep.ac.cn/pub/GECAM/GRBList/HEBS-GRB221009A.png}) and AGILE (\url{http://www.agilescienceapp.it/notices/GRB221009A_AGILE_RM.png}) satellites are used. In addition, Table \ref{tab:Table2} shows the energy range and mean energy in (MeV) with the mass attenuation coefficient of the mean energy used to calculate the attenuation of photon counts in Table \ref{tab:Table1}.

\begin{table}[htbp]
\centering
\begin{small}
\begin{tabular}{|c|c|c|c|c|c|}
\hline
\
\textbf{Altitude}& \multicolumn{3}{|c|}{\textbf{Fermi-GBM}}&\multicolumn{1}{|c|}{\textbf{SATech-01 HEBS}} &\multicolumn{1}{|c|}{\textbf{AGILE}} \\ 
(km) & \multicolumn{3}{|c|}{(535 km)}&\multicolumn{1}{|c|}{(535 km)} &\multicolumn{1}{|c|}{(500 km)} \\ 
\hline
&(0.01-0.044)&(0.044-0.3)&(0.3-1)&(0.4-6)&(0.4-100)\\
&MeV&MeV&MeV&MeV &MeV\\
\hline
\rowcolor{pink} {500} & 35000 & 75000 & 35000 & 35000 & 130000\\
\hline
50 & 22427.6 & 62873 & 31156.9 & 33634.3 & 126999\\
\hline
30 & 22.78 & 4095.24 & 5143.93 & 18158.8 & 88459.9\\
\hline
20 & 5.5e-08 & 1.58 & 28.81 & 3079.9 & 31236.9 \\
\hline
10 & 1.29e-48 & 1.24e-16 & 6.9e-10 & 0.7 & 230.53 \\
\hline
6 & 1.4e-78 & 1.7e-28 & 1e-17 & 0.001 & 6.18 \\
\hline
4 & 2.3e-99 & 9.7e-37 & 3.8e-23 & 2.1e-05 & 0.5 \\
\hline
0 & 3.4e-156 & 2.94e-59 & 5.3e-38 & 1.72e-10 & 0.00052 \\
\hline
\end{tabular}
\end{small}
\caption{Attenuation of photons from Low Earth Orbit to Earth surface during GRB 221009A}
\label{tab:Table1}
\end{table}

\begin{table}[htbp]
\centering
\begin{small}
\begin{tabular}{|c|c|c|}
\hline
\rowcolor{gray!40}
 \textbf{Energy range}& \textbf{Mean Energy} & \textbf{Mass attenuation}\\ 
 \rowcolor{gray!40}
 \textbf{} & & \textbf{coefficient ($\frac{\mu}{\rho}$)}\\
  \rowcolor{gray!40}
 \textbf{} & & \textbf{of mean energy}\\
 \rowcolor{gray!40}
\textbf(MeV) & \textbf(MeV) &  \textbf($m^{2}/kg$)\\ 
\hline
\ 0.01-0.044 & 0.03 & 0.03076 \\ 
\hline
\ 0.044-0.3 & 0.2 & 0.01219 \\ 
\hline
\ 0.3-1 & 0.6 & 0.008039\\
\hline
\ 0.4-6 & 5 & 0.002751\\
\hline
\ 0.4-100 & 50 & 0.001614\\
\hline
\end{tabular}
\end{small}
\caption{The mass attenuation coefficient ($\mu/\rho$) with the respective energy values; the complete coefficient list is available at \url{https://rais.ornl.gov/tools/Lin_Coefficients.pdf}}
\label{tab:Table2}
\end{table}

For energies (0.01 -- 100) MeV, equation \ref{eq3} results in essentially zero integral counts on the ground (Table \ref{tab:Table1}). Around 4 km, the number of photons diminishes to a very small fractional value, meaning all photons of these energies vanish. Thus, even if the NaI (Tl), or any other gamma-ray detector sensitive to energies $<$10 MeV, were placed at an altitude similar to that of LHAASO, it would not have detected any photons originating from this GRB. Or in other words, this GRB does not have high enough fluence of MeV photons. Another possibility of detecting MeV photons arises from the electromagnetic cascading of TeV photons via pair production and bremsstrahlung. However, while propagating, as the electrons and positrons of the shower approach the critical energy, $\epsilon_0$$\sim$ 80 MeV (in the air), the ionization becomes more important than bremsstrahlung, and the number of particles in the shower decreases. For primary photons of energies below 10 TeV (i.e., $10^{13}$ eV), the secondary photons in the shower die out before reaching the ground \citep{rossi1941cosmic,hillas1996differences,hoffman1999gamma}. Only for high-energy primary photons ($>$ 10 TeV), the shower reaches the ground, however, the shower development stops once the secondaries’ energies reach below the critical energy of $\sim$ 80 MeV. These secondary photons may still undergo Compton scattering; however, their energies will still be well beyond 10 MeV, and hence will not fall in the detection range of our NaI (Tl) detector. Thus not all the photons produced in the electromagnetic showers can reach the ground, and those that do, their energies are beyond the detection capability of the detector setup used in the present study. Unlike the NaI (Tl) detector, if the LHAASO setup was at our location, it could detect these very high-energy photons.

Further, this raises the question ``how much more intense a GRB needs to be observed from the ground in energies $<$10 MeV?''. Thus, the likelihood of detection of photons $<$10 MeV is investigated for a hypothetical GRB even more intense than the GRB 221009A. Further, the parameters of such hypothetical GRB are estimated when ground observation of gamma-rays ($<$10 MeV) will be possible.  

The goal here is to estimate the $E_{iso}$ and fluence of a hypothetical GRB, which requires the photon flux at LEO.
Also, it is pertinent to determine if photons ($<$10 MeV) from this hypothetical GRB will reach the ground. For this purpose, the Beer-Lambert attenuation equation can be used in the following manner: first, we assume that a certain number of photons, e.g., 100 (of energy 5 MeV), reach the ground during the hypothetical GRB. Then, the inverse equation of equation \ref{eq3}, 
\begin{equation} \tag{4}\label{eq4}
    	I = I{_{o}}.e^{\displaystyle(\frac{\mu}{\rho}).\rho . x},
\end{equation}
where ${x}$ varies from 0 to 500 km, and $I_o$ is the number of photons at 0 km, i.e., 100. Equation \ref{eq4} thus gives the back-calculated, corresponding number of photons, i.e., $~N$(5 MeV) at LEO. The different values of $I$ at increasing altitudes are shown in Table \ref{tab:Table3}. However, to calculate $~N(E)$ for all energies in the (0.1 -- 10) MeV range, there will be obstacles such as limitation on finer steps of energies and uncertainties in ($\mu/\rho$) for these energies. Thus, such calculation is only used for cross-verification.

To estimate $~N(E)$ at LEO in the (0.1 -- 10) MeV energy range, the above calculation can be used in combination with the Band function, which represents the distribution of gamma counts as energy spectrum for GRBs \cite{band1993batse}. 

\begin{equation} \tag{5}\label{eq5}  
N(E) =  A { \left( \frac{E}{100~keV} \right)}^\alpha exp { \left( \frac{-E}{E_{0}} \right)},~for~ E< (\alpha - \beta)E_{0},~~~~~~~~~~~~~~~~~~~~~~~~~~~~~~~~~~~~~~~~ 
\end{equation}

\begin{equation}\tag{6}\label{eq6}
N(E) =  A {{\left[{\frac{(\alpha - \beta)E_0}{100~keV}}\right]}^{\left(\alpha - \beta \right)}}exp{\left(\alpha - \beta \right)}{\left( \frac{E}{100~keV}\right)^\beta},~for~E \ge (\alpha - \beta)E_0,~~~~~~~~~~~~~~~~~~ 
\end{equation}
where $~A$ is a normalising parameter, and the two spectral parts are separated by the break energy $E_{0}$. From hundreds of previous GRBs, standard values of $\alpha$ and $\beta$ are known to be approximately -1 and -2, respectively \citep{band1993batse}. 

 The resulted value $~N$(5 MeV) from equation \ref{eq4} is thus compared with that obtained from equations \ref{eq5} and \ref{eq6}, and this gives us such a normalising factor that will allow a fraction of the total $~N(E)$, i.e., counts of all energies in (0.1 -- 10) MeV range that is obtained from the Band function, to reach the ground. This $~N(E)$ is then used to calculate the fluence of the hypothetical GRB. Using $S = \int{N(E).EdE}$, fluence for a burst duration of 10 s is calculated as $\sim$ 1.15$\times10^{13}~erg~cm^{-2}$ ($\approxeq$ 1.15$\times10^7~kJ~m^{-2}$), which is $\sim$ 10$^{14}$ times the fluence of GRB 221009A. Hypothetical GRB possibly responsible for mass extinction as discussed in \cite{melott2004did,thomas2005terrestrial,melott2005climatic} has fluence of 100 $kJ~m^{-2}$, i.e., $10^8~erg~cm^{-2}$.

\begin{table}[htbp]
\centering
\begin{small}
\begin{tabular}{|c|c|}
\hline
 \textbf{Altitute}& \textbf{Photon Counts (Hypothetical GRB)}\\ 
\textbf(km) & (0.1-10) MeV \\ 
\hline
\rowcolor{pink} 0&100 \\
\hline
4&1.21e+07 \\
\hline
6&8.72e+08\\
\hline
10&4.17e+11  \\
\hline
20&1.79e+15 \\
\hline
30&1.06e+16\\
\hline
50&1.96e+16 \\
\hline
500& 2.04e+16\\
\hline
\end{tabular}
\end{small}
\caption{Photons counts from Earth surface to LEO for hypothetical GRB, by considering attenuation mass coefficient ($\mu/\rho$) = 0.002751 ($m^{2}/kg$) for mean energy $~E$ = 5 MeV}
\label{tab:Table3}
\end{table}

If the hypothetical GRB is occurring at 2 $~kpc$ from Earth, like in \cite{thomas2005terrestrial}, isotropic energy will be $E_{iso}$ = 5.53$\times$$10^{57}~erg$. However, $E_{iso}$ assumed in Thomas et al. (2005) \cite{thomas2005terrestrial} is 4.79$\times$$10^{52}~erg$, the average $E_{iso}$ of observed GRBs in record is $10^{54}~erg$, and the maximum is that of GRB 221009A, $E_{iso}$ = $1.2\times10^{55}~erg$  \citep{burns2023grb}. Thus, if the total energy condition is obeyed, for $E_{iso} = 10^{55}~erg$, and calculated fluence of $\sim$ 1.15$\times10^{13}~erg~cm^{-2}$, the distance of such GRB comes out to be $\sim$ 0.1 $~kpc$ ($\sim$ 326$~ly$).

Thus, if a GRB of isotropic energy $10^{55}~erg$ occurs at $\sim$ 0.1 $~kpc$ from the Earth, gamma-rays of energy (0.1 -- 10) MeV will reach the ground. However, the probability of occurrence of such GRB is extremely low.

\section{Conclusion}
The GRB 221009A was a long-lasting and bright gamma-ray burst (GRB) observed by various space and ground-based instruments. Though it is the brightest of all time, no significant change over the background secondary cosmic rays (SCR) photons is observed in NaI (Tl) detectors on the ground during GRB 221009A. Even though, theoretically, the NaI (Tl) detector employed in our study is capable of measuring both GRB and SCR photons, GRB origin photons of energies below 10 MeV are not observed due to differences in the attenuation and propagation/generation mechanisms of these photons in the Earth's atmosphere. Estimation of attenuation of GRB originated gamma-rays in the atmosphere suggests that photons below 10 MeV vanish upwards of 4 km altitude. We conclude that the low energy ($<$10 MeV) gamma-rays originating from GRBs do not reach the ground, except in extremely rare circumstances of a bright GRB occurring nearby, e.g., a GRB of isotropic energy $E_{iso} = 10^{55}~erg$ and fluence of the order of $\sim$ 10$^{13}~erg~cm^{-2}$, occurring within 0.1 $~kpc$ from Earth.

\acknowledgments

The experimental set-up at Tirunelveli is operated by the Indian Institute of Geomagnetism (IIG). The data used in this work are available at server IP $59.185.241.6$ with FTP access using `anonymous' option for login; data is provided in `Data/NaI$\textunderscore$SCR$\textunderscore$data$\textunderscore$Tirunelveli' folder. However, continuous data of NaI (Tl) detectors is available on request through permission from the Director, IIG. This work is supported by the Department of Science and Technology, Government of India.

\bibliographystyle{unsrt}
\bibliography{GRB221009A_without_highlight}

\end{document}